# A perspective magnetic bed comprising Gd alloy multi-microwires for energy-efficient magnetic refrigeration

Hongxian Shen[1,2], Lin Luo[1], Sida Jiang[3,*], Jingshun Liu[5], Yanfen Liu[5], Yongjiang Huang[1], Lunyong Zhang[1], Jianfei Sun[1,*], Hillary Belliveau[2], and Manh-Huong Phan[2,*]

[1] School of Materials Science and Engineering, Harbin Institute of Technology, Harbin 150001, P. R. China

[2] Department of Physics, University of South Florida, Tampa, Florida 33620, USA

[3] Space Environment Simulation Research Infrastructure, Harbin Institute of Technology, Harbin 150001, China

[4] School of Materials Science and Engineering, Inner Mongolia University of Technology, No. 49 Aimin Street, Hohhot 010051, P. R. China

[5] Physics Department, Qiqihar University, Qiqihar 161006, Heilongjiang, P. R. China

We have designed a new magnetic bed structure with desirable table-like magnetocaloric effect (MCE) by using three kinds of soft ferromagnetic Gd-Al-Co microwire arrays with different Curie temperatures ($T_C$). The $T_C$ interval of these three wires is ~10 K and the designed new structure named Sample A. This sample shows a smooth table-like magnetic entropy change ($\Delta S_M$) at high applied field change ($\mu_0 \Delta H$=5 T) ranging from ~92 K to ~107 K. The maximum entropy change ($-\Delta S_M^{max}$) and refrigerant capacity (RC) for Sample A at $\mu_0 \Delta H$=5 T are calculated to be ~9.42 J·kg$^{-1}$·K$^{-1}$ and ~676 J·kg$^{-1}$. The calculated curves of $-\Delta S_M(T)$ and the corresponding experimental data match well with each other, suggesting that the desirable magnetocaloric properties of the microwire arrays can be designed. Simulation shows that the

RC values of the designed systems increase when increasing the interval of $T_C$. The table-like MCE and the enhanced heat-transfer efficiency due to the enhanced surface areas of the microwires make this newly designed magnetic bed very promising for use in energy-efficient magnetic refrigerators.

*Corresponding authors: jiangsida@hit.edu.cn (S.D. Jiang,) jfsun_hit@263.net (J.F. Sun), phanm@usf.edu (M.H. Phan)

# 1. Introduction

Magnetic refrigeration based on the magnetocaloric effect (MCE) is a promising alternative to conventional vapor compression refrigeration techniques due to its higher efficiency, compact design, lack of moving parts, and the use of non-polluting environmental materials [1-5]. The heat transfer during the MCE process is dependent upon the entropy change in the magnetic refrigerant element. Therefore, a material system exhibiting a large entropy change is highly valuable to magnetic refrigeration technologies. An entropy change can be induced in magnetic refrigerant materials by either magnetic or structural phase transitions [1,5]. It is widely known that while first-order magnetic transition materials (FOMT) possess larger values of magnetic entropy change ($\Delta S_M$), they are characteristically limited to narrower temperature ranges [6]. This is the limiting factor of the operating temperature of this type of magnetic material. However, second-order magnetic transition materials (SOMT) have lower values of $\Delta S_M$, which extends through a broader temperature range. Because of the broader operating temperature range as well as negligible thermal hysteretic losses, the refrigerant capacity (RC) of SOMT materials is generally larger than for FOMT materials. The standard SOMT material for operating temperatures considerably higher than cryogenic temperatures is gadolinium (Gd). Since Gd is costly, alloying it with other non-rare-earth elements, such as $Gd_5(Si_xGe_{1-x})_4$ compounds [1,5], has proven to be useful for improving the MCE while reducing the material cost considerably.

Kuzmin theoretically showed that reducing the dimensions of a magnetic refrigerator could increase the cooling power of the device by increasing the operating frequency [7]. In this work, the author indicated that shaping magnetic refrigerants in the form of spherical or irregular particles is inefficient, due to their high losses on viscous resistance and demagnetization, and suggests the microwire geometry to be used instead. Mechanical instability of the refrigerant can result in a significant loss of heat throughout due to unequal distribution of flow. In this context, the use of a bundle of magnetocaloric wires (e.g. Gd wires) has been proposed to be more desirable because this configuration enables higher mechanical stability and lower porosity while simultaneously increasing the operating frequency of the magnetic refrigerant material. Vuarnoz and Kawanami have theoretically presented that a magnetic bed made of an array of Gd wires produces a greater temperature span between its ends, which results in a higher cooling load at a higher efficiency, as compared to a magnetic bed made of Gd particles [8]. Unlike their bulk counterparts, the use of the wires with increased surface areas also allows for a higher heat transfer between the magnetic refrigerant and surrounding liquid [7-9]. These theoretical studies have opened up new areas for magnetic refrigeration device and material design. Experimental studies in Gd-alloy microwires have already illustrated excellent magnetocaloric properties in Gd-alloy microwires [10-22].

For magnetic cooling applications above 20 K, the regenerative Ericsson cycle is theoretically ideal for the magnetic cooling systems due to its high working efficiency and broad temperature range [23]. Thus, magnetic refrigerants showing broad working temperature ranges are required which means the materials has table-like MCE near the maximum entropy change ($\Delta S_M^{max}$). Typically, the shapes of the magnetic entropy change as a function of temperature for alloys or elementary substances are shape peaks [24,25] or broad triangular peaks [5]. Some new structures have been designed and there are two distinctive methods. The first method is fabricating alloys with multiple phases [26, 27]. Materials with different phases usually show different MCE properties such as different Curie temperatures ($T_C$) and the MCE of the alloys are combinations of these respective phases. The multiphases are commonly formed during the fabrication process [23] or obtained through thermal annealing. Sometimes it is difficult to control the amount of each individual phase or obtain the required phase. Another method is designing composites or structures by utilizing alloys with different compositions and MCE properties [28,29], such as sintering magnetic materials with different compositions [30,31] and fabricating multi-layers structures with different ribbons [32,33]. The MCE properties of the composites are easy to control and designed based on the MCE of different alloys and some multi-layers structures are good for heat exchange in the cooling systems. Table 1 shows tunable MCE and Curie temperature values in microwires with varied compositions. This adds more interest to study devices made of Gd-alloy in the microwire geometry.

In this work, we have used the second method to explore a new design for novel magnetic cooling systems (Fig. 1a,b) for obtaining table-like MCE by using melt-extracted Gd-Al-Co microwire arrays with different compositions and MCE properties.

## 2. A magnetic bed made of laminate-arrayed Gd alloy microwires: Design, fabrication, and characterization

The Gd-Al-Co microwires were fabricated through the melt-extracted method and the process can be found in our previous reports [13-16]. Fig. 1c is the SEM image of melt-extracted Gd-Al-Co microwires which are uniform and show diameters of ~30 μm. The Gd-Al-Co microwires exhibit amorphous structures due to the quench rate of melt-extraction and sometimes a few nano-crystal phases appear on the amorphous matrix. The MCE properties of these Gd-based microwires can be tuned by adjusting the composition. Moreover, the microwires show excellent mechanical properties and MCE properties, which promote enough mechanical strength when applied in magnetic cooling system due to their amorphous nature [15]. Furthermore, the microwires show high adaptation of morphology and structure thus magnetic refrigerant with different porous structures can easily be designed by arranging these microwires, as shown in Fig. 1a. It is worth noting that the micro-size of these wires enhances the surface for the designed structures, which improved the efficiency of heat transfer significantly in the cooling system. These excellent magnetocaloric properties and mechanical performances make these melt-extracted Gd-Al-Co microwires as one of the best candidate materials in magnetic refrigeration designs.

We have chosen three kinds of Gd-Al-Co microwires with surrounding maximum entropy changes but different $T_C$ temperatures for designing multi-microwires structures and obtaining table-like MCE. The MCE performances of these kinds of Gd-Al-Co microwires are shown in

Table 2 and the $T_C$ interval is ~ 10 K [13,16]. We designed a specimen by using microwires arrays with $Gd_{50}Al_{30}Co_{20}$, $Gd_{50}Al_{25}Co_{25}$ and $Gd_{60}Al_{20}Co_{20}$ compositions according to Fig. 1a and all the wires of Sample A have equal weight fractions. The designed specimens as magnetic refrigerants are schematically shown in Fig. 1b.

For evaluating the magnetic transitions of Sample A formed by different kinds of microwires, magnetization dependence on temperatures (*M-T*) curves ranging from 20-200 K at applied field of 200 Oe were measured on a commercial Physical Property Measurement System from Quantum Design and the results are displayed in Fig. 2b. The *M-T* curves of the respective components we measured before are also shown in Fig. 2a for comparison. Notably, the *M-T* curve of Sample A shows a smooth ferromagnetic-paramagnetic (FM-PM) transition which exhibits the similar tendency with its components. In order to understand the transition behavior of Sample A, the differential curves (*dM/dT vs. T*) were calculated and shown in inset of Fig. 2b. We also have calculated the differential curves of its components and the results are plotted in inset of Fig. 2a. The differential curve of Sample A shows only single peak near the magnetic transition though they are designed by three kinds of microwires with different $T_C$ temperatures. As we know that the peak of *dM/dT vs. T* curve is usually defined as $T_C$ temperature, thus means the designed Sample A has its own $T_C$ temperature which is determined as ~100 K.

Based on thermodynamic Maxwell equation, the magnetic entropy change can be calculated as

$$\Delta S_M = \mu_0 \int_0^{H_{max}} \left( \frac{\partial M}{\partial T} \right)_H dH$$

(1)

where $\Delta S_M$ donates the magnetic entropy change, $M$, $H$ and $T$ is the magnetization, magnetic field and temperature, respectively, the $\Delta S_M$ results can be calculated through a series of isothermal magnetization (*M-H*) curves. Thus, the *M-H* data of Sample A were collected at applied field change ($\mu_0 \Delta H$) of 5 T and different temperatures ranging of 20-200 K with a temperature interval of 10 K and decrease to 5 K near transition temperatures then decreased to 1 K around transition temperatures. The measured results of *M-H* data for Sample A are plotted in Fig. 2c, then the calculated -$\Delta S_M$ (*T, H*) dependences of Sample A are plotted as three dimensional forms and shown in Fig. 2d, respectively. The maximum entropy change (-$\Delta S_M^{max}$) of Sample A at $\mu_0 \Delta H$ =5 T is ~9.42 J·kg$^{-1}$·K$^{-1}$ and the value is lower than those of all its components (~10.09 J·kg$^{-1}$·K$^{-1}$, ~10.3 J·kg$^{-1}$·K$^{-1}$ and ~10.11 J·kg$^{-1}$·K$^{-1}$ for $Gd_{50}Al_{30}Co_{20}$, $Gd_{55}Al_{25}Co_{25}$ and $Gd_{60}Al_{20}Co_{20}$ wires at $\mu_0 \Delta H$ =5 T, respectively). The decrease of -$\Delta S_M^{max}$ for Sample A is due to the large $T_C$ intervals among the Gd-Al-Co microwires.

Remarkably, Sample A clearly exhibits a table-like MCE behavior as shown in the 3D plot of Fig. 2d even at low applied field changes. For further to exploring the MCE of Sample A near the transition temperatures, we choose the -$\Delta S_M$(*T*) curve at $\mu_0 \Delta H$ =5 T which is plotted in Fig. 3b and the -$\Delta S_M$ *vs*. *T* curves at $\mu_0 \Delta H$ =5 T of its respective components are also displayed in Fig. 3a for comparison. Clearly the table-like MCE is obtained in Sample A as shown in inset of Fig. 3b. The table-like MCE for Sample A shows a little fluctuation at its transition region and the width of the table is ~15 K ranging from ~92 K to 107 K. The fluctuation may be caused due to the relative larger $T_C$ intervals and its $Gd_{60}Al_{20}Co_{20}$ component, which shows fluctuation at the transition region as shown in Fig. 3a.

The wires are mixed mechanically and assumed without any interactions among

themselves, and then we can calculate the magnetic entropy changes of Sample A based on a simple equation:

$$\Delta S_M^{Design} = \alpha \Delta S_M^1 + \beta \Delta S_M^2 + \gamma \Delta S_M^3 + \cdots\cdots + \omega \Delta S_M^n \tag{2}$$

where $\Delta S_M^{Design}$ is the magnetic entropy change of designed structures, $\alpha$, $\beta$, $\gamma$ and $\Delta S_M^1$, $\Delta S_M^2$, $\Delta S_M^3$ are the weight fractions and magnetic entropy changes of the different microwires. The calculated result of Sample A is plotted in Fig. 3b which shows only a slight deviation when $T > T_C$. This means the calculated data based on Eq. (2) fits with the experimental data very well, demonstrating that Eq. (2) can be applied for designing new structures by using soft ferromagnetic Gd-Al-Co multi-microwires which are mixed mechanically and with insignificant interactions among themselves.

For exploring the difference of MCE behaviors between the designed structure and the respective Gd-Al-Co wires, we have used universal master curves to evaluate the changes that can be established by collapsing all $\Delta S_m$ vs. $T$ curves at their external fields through $-\Delta S_m$ normalized to $-\Delta S_m^{max}$. Their temperature axes also need to be rescaled by equation [34]:

$$\theta = \begin{cases} -(T-T_C)/(T_{r2}-T_C) & T \leq T_C \\ (T-T_C)/(T_{r1}-T_C) & T \geq T_C \end{cases} \tag{3}$$

where $T_{r1}$ and $T_{r2}$ are two reference temperatures above and below $T_C$ which should be satisfied the relation:

$$-\Delta S_m(T_{r1}) = -\Delta S_m(T_{r2}) = f \times (-\Delta S_m^{max}) \tag{4}$$

For the present work, the value of $f$ is determined as 0.5. All the applied Gd-Al-Co microwires show second-order magnetic transition characters, thus all the universal master curves at different external fields can be fitted very well [16]. Herein, we chose the universal master curves at $\mu_0 \Delta H = 5$ T for comparison. The calculated universal master curves of Sample A and its Gd-Al-Co components are displayed in Fig. 3c. The universal master curves of its components are fitted very well, which explain that all these Gd-Al-Co wires show almost same MCE and magnetic transition behaviors with ~10 K $T_C$ intervals. But the universal master curve of Sample B shows clear differences with the Gd-Al-Co microwires when $T>T_C$. That means the MCE behavior of the designed structure changed when its components with ~10 K $T_C$ intervals were used.

For comparing the cooling efficiency between the designed sample (Sample A) and its components, the RC at $\mu_0\Delta H$=5 T for Sample A is also calculated as ~676 J·kg$^{-1}$ according to the equation:

$$RC = \int_{T_{cold}}^{T_{hot}} -\Delta S_M(T) dT \qquad (5)$$

where $T_{cold}$ and $T_{hot}$ are the temperatures at the full width at half-maximum (FWHM) of -$\Delta S_M(T)$ curves. The RC value of Sample A is a little higher than ~659 J·kg$^{-1}$ which calculated by the average RC values of its three components at $\mu_0\Delta H$ = 5 T. The RC values of Sample A are not increased very much compared with their components, due to the smaller $T_C$ intervals of these microwires.

For exploring the changing of MCE and RC properties for the designed structures with different $T_C$ intervals, we have applied virtualized data and calculated the -$\Delta S_M$ vs. T curves at $\mu_0\Delta H$ = 5 T based on Eq. (2). The virtualized data are based on our experimental data and we just changed the $T_C$ temperatures for obtaining new data by increasing or decreasing same temperature values. It has been shown that the RC value of a designed structure increases compared with its two components when its two components have relatively larger $T_C$ intervals [5,29,30]. Thus, we first have designed five samples (named S-1, S-2, S-3, S-4 and S-5) by two components with different $T_C$ intervals. We have used the experimental data of $Gd_{50}Al_{30}Co_{20}$ microwires as one basic composition (named C-Basic) and $Gd_{50}Al_{25}Co_{25}$ microwires as other varying compositions (C-1, C-2, C-3, C-4 and C-5). The $T_C$ intervals between varying compositions (C-1, C-2, C-3, C-4, C-5) and C-Basic are ~10 K, 20 K, 30 K, 40 K and 50 K, respectively, as shown in Fig. 4a. We assumed there are no magnetic interactions among the wires and all the components had same weight fractions. Then the $-\Delta S_M$ vs. $T$ curves at $\mu_0\Delta H$=5 T of the designed samples can be calculated based on Eq. (2) and the results are displayed in Fig. 4a which clearly shows that the designed samples exhibit more and more obviously table-like MCE with the increasing $T_C$ intervals. Additionally, the values of $-\Delta S_M^{max}$ decrease when the $T_C$ intervals increase. The RC values of these designed samples at $\mu_0\Delta H$=5 T are calculated based on Eq. (5) and plotted in Fig. 4c, the RC values of their components (C-1, C-2, C-3, C-4 and C-5 have same RC values) at $\mu_0\Delta H$=5 T are also plotted in Fig. 4c for comparison. The RC value of S-1 is lower than that of C-Basic when $T_C$ interval is ~10 K while RC values of other samples are higher compared to both the components. Remarkably, the RC values of the designed samples increase with increasing the $T_C$ interval, which means enhanced RC values for designed systems can be obtained by using the Gd-based wires with large $T_C$ intervals in a range of temperature.

In order to study if the designed systems with three components showing same tendency as structures with two components, we have simulated the design of another five samples (named S'-1, S'-2, S'-3, S'-4 and S'-5) by three components with different $T_C$ intervals. Here we have applied the experimental data of $Gd_{50}Al_{25}Co_{25}$ microwires as the basic composition (named C'-Basic), $Gd_{50}Al_{30}Co_{20}$ microwires as one varying compositions (C 1-1, C 1-2, C 1-

3, C 1-4 and C 1-5) and $Gd_{60}Al_{20}Co_{20}$ microwires as another varying composition (C 2-1, C 2-2, C 2-3, C 2-4 and C 2-5). The $T_C$ intervals between C'-Basic and C 1-1, C 1-2, C 1-3, C 1-4, C 1-5 are ~10 K, 20 K, 30 K, 40 K and 50 K, respectively, and between C'-Basic and C 2-1, C 2-2, C 2-3, C 2-4, C 2-5 are ~-10 K, -20 K, -30 K, -40 K and -50K, respectively, as displayed in Fig. 4b. According to Eq. (2), we have calculated the $-\Delta S_M$ vs. $T$ curves at $\mu_0\Delta H$=5 T of the designed systems and also plotted them in Fig. 4b. The designed systems with three components exhibit the same tendency as those with two components, which show more and more obviously table-like MCE and the values of $-\Delta S_M^{max}$ decrease when the $T_C$ intervals increase. Fig. 4d shows the calculated RC values of these designed systems and their components (C 1-1, C 1-2, C 1-3, C 1-4 and C 1-5 have same RC values, also for C 2-1, C 2-2, C 2-3, C 2-4 and C 2-5) at $\mu_0\Delta H$=5 T. Notably, the RC values of the designed systems with three components also exhibit the same trend as those with two components which increase as the $T_C$ intervals increase. The enhanced RC values can also be obtained by designing systems using three components with large $T_C$ intervals.

## 3. Conclusion

In summary, we have successfully obtained desirable table-like MCEs near the transition region in the newly designed Sample A using soft ferromagnetic Gd-Al-Co microwires arrays. The values of $-\Delta S_M^{max}$ and RC at $\mu_0\Delta H$=5 T for Sample A are ~9.42 J·kg$^{-1}$·K$^{-1}$ and ~676 J·kg$^{-1}$. The universal master curves clearly show the differences between the designed system and Gd-Al-Co wires. Furthermore, the calculated $-\Delta S_M$ vs. $T$ curves based on Eq. (2) fitted the experimental data very well, suggesting that the magnetic refrigerants with table-like MCEs applied at liquid nitrogen region for novel magnetic cooling systems can be designed by melt-extracted Gd-Al-Co microwires according to this equation. The enhanced RC values can be obtained by designing the systems using the wires with large $T_C$ intervals. In future work, we will focus on designing the structures using more (>3) Gd-based microwires components.

**Acknowledgments**

The work was partially supported by the National Natural Science Foundation of China (NSFC, Nos. 51801044, 51901057, 51827801). M.H.P acknowledge support from the U.S. Department of Energy, Office of Basic Energy Sciences, Division of Materials Sciences and Engineering under Award No. DE-FG02-07ER 46438.

**Table 1.** Magnetocaloric properties of the melt-extracted Gd-Al-Co microwires.

|  | Composition | $T_C$ /(K) | $\mu_0\Delta H$=5 T | | |
|---|---|---|---|---|---|
|  |  |  | $-\Delta S_M$ /(J·kg$^{-1}$·K$^{-1}$) | RC /(J·kg$^{-1}$) | RCP /(J·kg$^{-1}$) |
|  | Gd$_{50}$Al$_{30}$Co$_{20}$ | 86 | 10.09 | 672 | 861 |
| **Sample A** | Gd$_{50}$Al$_{25}$Co$_{25}$ | 97 | 10.30 | 622 | 833 |
|  | Gd$_{60}$Al$_{20}$Co$_{20}$ | 109 | 10.11 | 681 | 915 |

**Figure captions**

**Fig. 1.** Magnetic refrigerant designed by arranging soft magnetic multi-microwires with a) single-component and b) with three kinds of components. c) SEM image of melt-extracted Gd-Al-Co microwires.

**Fig. 2.** Magnetization dependence on temperature curves of a) all applied Gd-Al-Co microwires and b) designed Sample A at external field of 200 Oe. Insets are their corresponding differential curves. c) Isothermal magnetization curves (*M-H*) of Sample A at different temperatures and magnetic field changes and d) is its corresponding calculated magnetic entropy changes dependence on temperatures and field changes.

**Fig. 3.** a) shows the -$\Delta S_M$ *vs*. *T* curves at $\mu_0 \Delta H$ =5 T of its components for Sample A and b) are the -$\Delta S_M$ *vs*. *T* curves at $\mu_0 \Delta H$ =5 T of Sample A and its calculated results based on Eq. 1, respectively. c) and d) are the universal master curves and RC values at $\mu_0 \Delta H$ =5 T of Sample A and its Gd-Al-Co components, respectively.

**Fig. 4.** Simulated -$\Delta S_M$ *vs*. *T* curves at $\mu_0 \Delta H$=5 T of the designed systems and their components for a) two components systems and b) three components systems. RC values of a) Sample A and b) Sample B and its Gd-Al-Co components

**Figure 1**

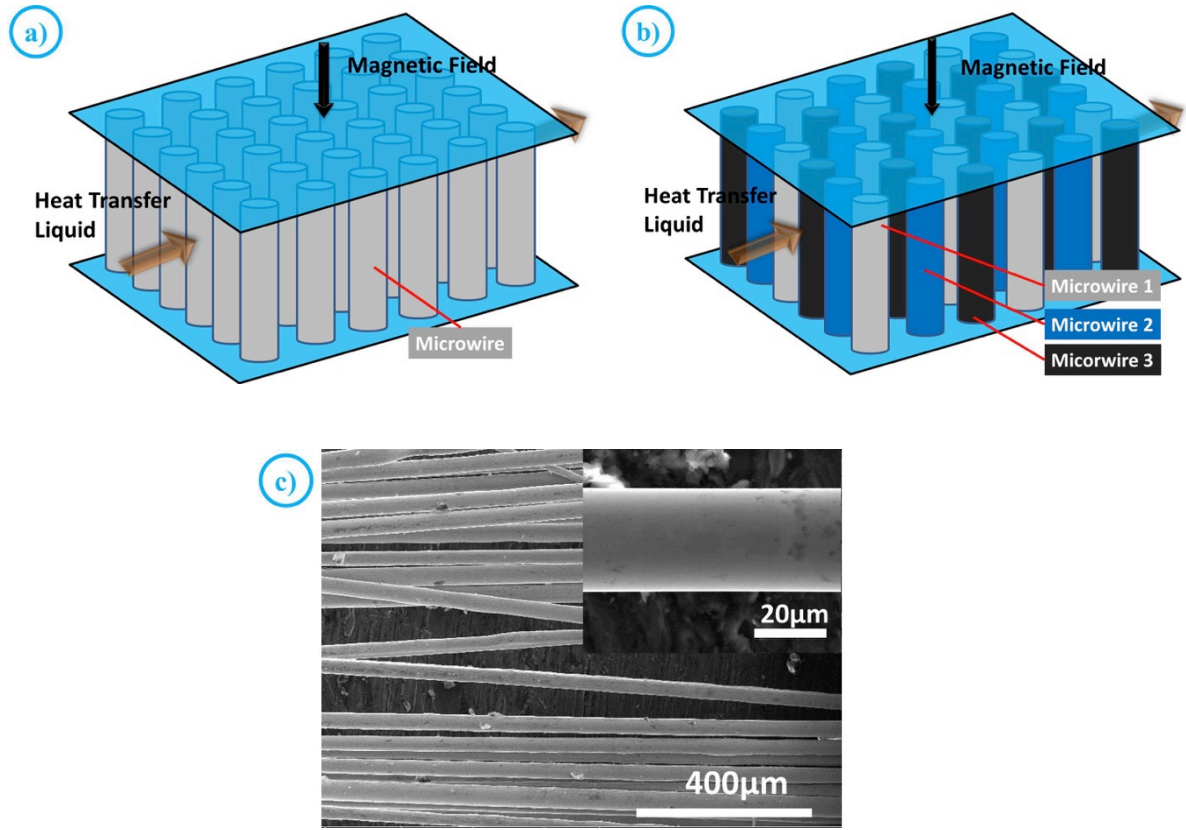

**Figure 2**

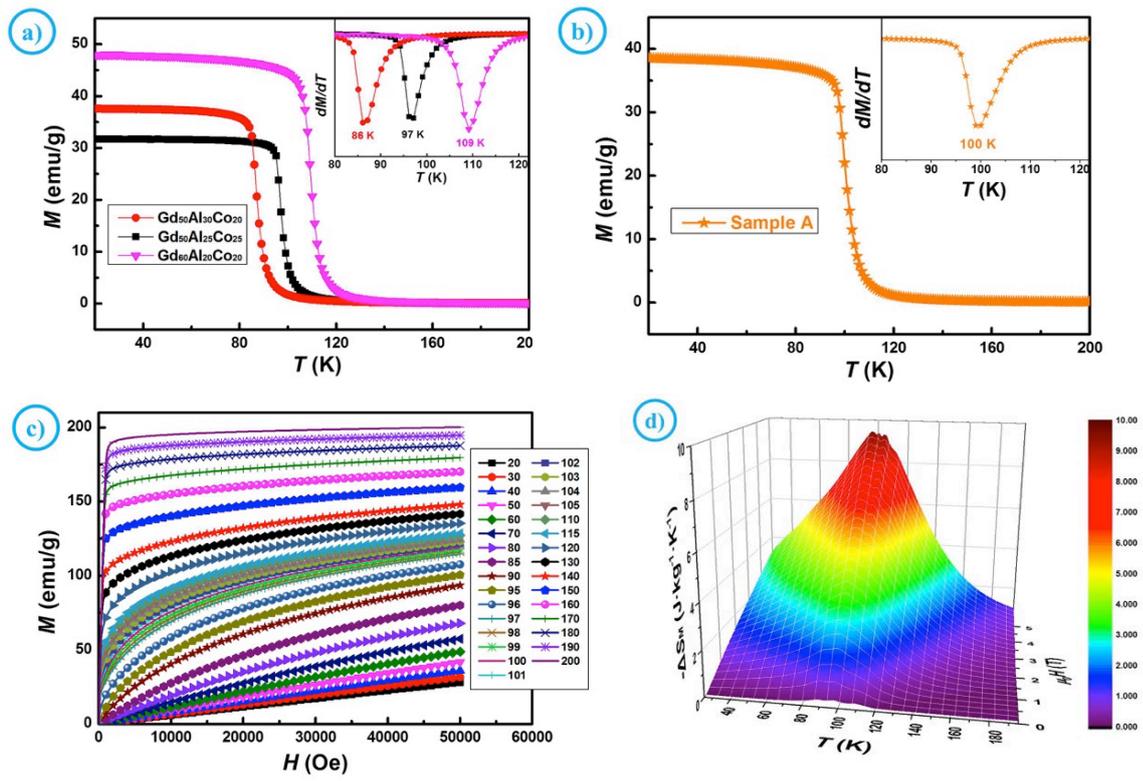

**Figure 3**

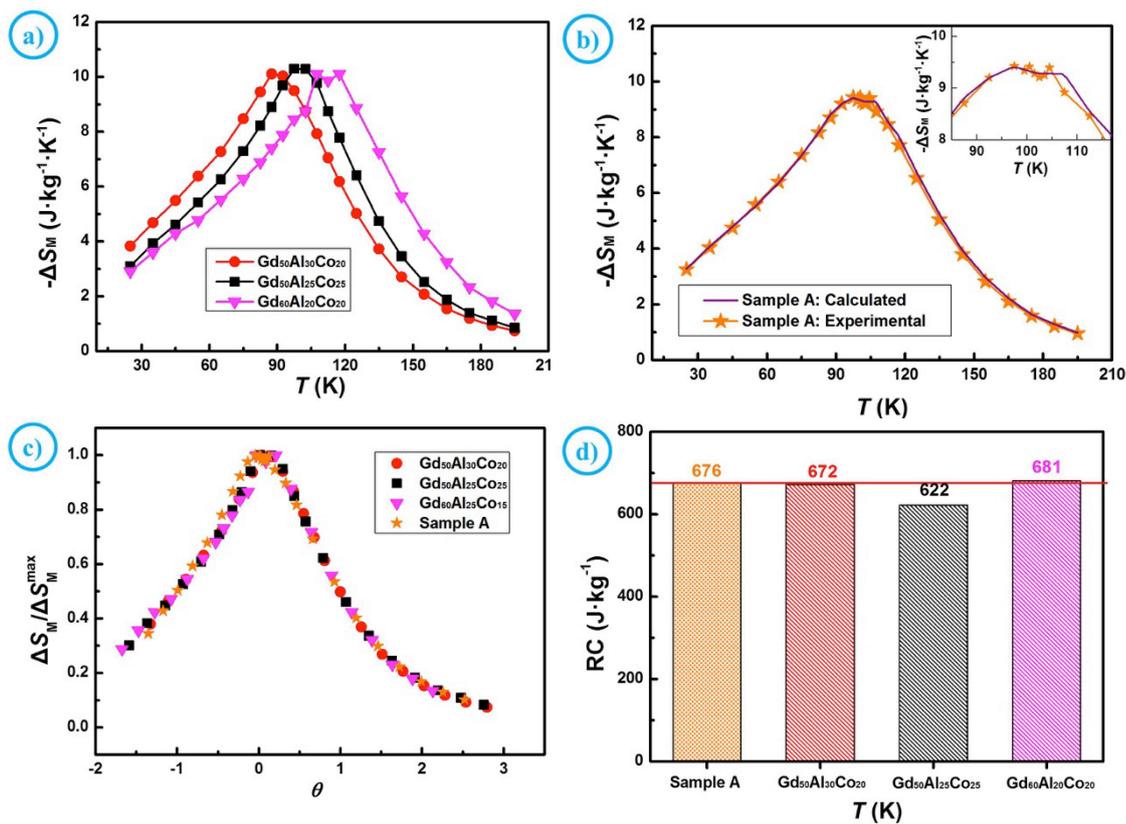

**Figure 4**

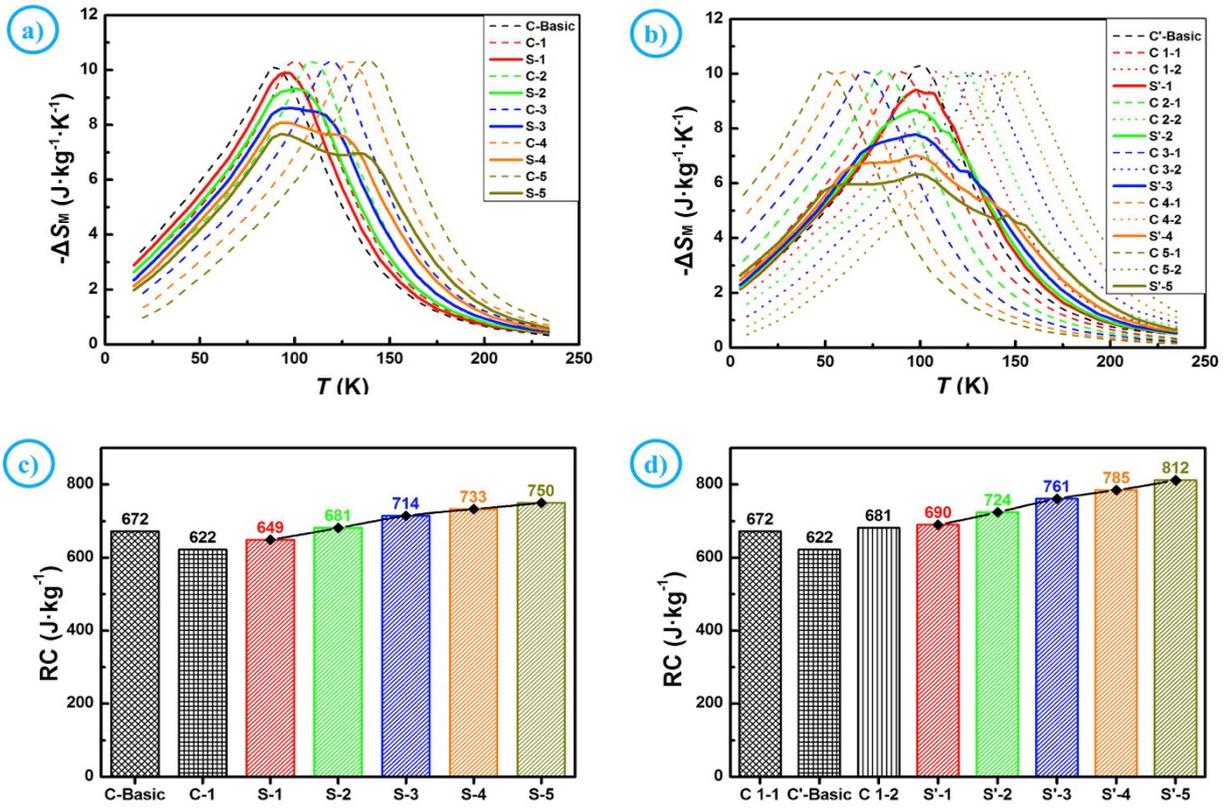